\documentclass[useAMS,usenatbib,usegraphicx]{mn2e}

\usepackage{times,epsfig}

\def\gsim{\;\rlap{\lower 2.5pt
 \hbox{$\sim$}}\raise 1.5pt\hbox{$>$}\;}
\def\lsim{\;\rlap{\lower 2.5pt
   \hbox{$\sim$}}\raise 1.5pt\hbox{$<$}\;}
\def\micron{~$\mu\textrm{m}$ }
\def\micronend{$\mu\textrm{m}$}

\title[A physical model for the diffuse cosmic infrared background]
       {A physical model for the origin of the diffuse cosmic infrared background and the opacity of the Universe to very high energy $\gamma$-rays}
\author[Younger \& Hopkins]{Joshua D. Younger$^{\! 1,2}$\thanks{jyounger@ias.edu} \& Philip F. Hopkins$^{\! 3}$ \\
	   \vspace*{1mm}\\
             $^1$ Hubble Fellow \\
             $^2$ School of Natural Sciences, Institute for Advanced Study, Einstein Drive, Princeton, NJ 08540, USA \\
             $^3$ Department of Astronomy and Theoretical Astrophysics Center, University of California Berkeley, Berkeley, CA 94720, USA
             }

\date{\fbox{\sc Draft dated: \today\ }}
\pagerange{000--000}

\begin{document}

\maketitle 

\begin{abstract}

We present a physical model for origin of the cosmic diffuse infrared background (CDIRB).  By utilizing the observed stellar mass function and its evolution as input to a semi-empirical model of galaxy formation, we isolate the physics driving diffuse IR emission.  The model includes contributions from three primary sources of IR emission: steady-state star formation owing to isolated disk galaxies, interaction-driven bursts of star formation owing to close encounters and mergers, and obscured active galactic nuclei (AGN).  We find that most of the CDIRB is produced by equal contributions from objects at $z\sim 0.5-1$ and $z\gsim 1$, as suggested by recent observations.  Of those sources, the vast majority of the emission originates in systems with low to moderate IR luminosities ($L_{IR} \lsim 10^{12}$ $L_\odot$); the most luminous objects contribute significant flux only at high-redshifts ($z\gsim 2$).  All star formation in ongoing mergers accounts for $\lsim 10\%$ of the total at all wavelengths and redshifts, while emission directly attributable to the interaction-driven burst itself accounts for $\lsim 5\%$.  We furthermore find that obscured AGN contribute $\lsim 1-2\%$ of the CDIRB at all wavelengths and redshifts, with a strong upper limit of less than $4\%$ of the total emission.  Finally, since electron-positron pair production interactions with the CDIRB represent the primary source of opacity to very high energy (VHE: $E_\gamma \gsim 1$ TeV) $\gamma$-rays, the model provides predictions for the optical depth of the Universe to the most energetic photons.  We find that these predictions agree with observations of high-energy cutoffs at $\sim $ TeV energies in nearby blazars, and suggest that while the Universe is extremely optically thick at $\gsim 10$ TeV, the next generation of VHE $\gamma$-ray telescopes can reasonably expect detections from out to $\sim 50-150$ Mpc.

\end{abstract}


\section{Introduction}
\label{sec:intro}

Diffuse extragalactic background light (EBL) represents the sum total of all the photons produced by luminous matter over the lifetime of the Universe.  In the ultraviolet, optical, and near-infrared (IR) the EBL is directly attributable to star formation and active galactic nucleus (AGN) activity \citep[for reviews, see][and references therein]{tyson1990,tyson1995,henry1991,henry1999,leinert1998}.  However, in regions with significant dust opacity radiation generated by stars and AGN is reprocessed by dust into the IR as thermal emission \citep[see, e.g.][]{soifer1991,sanders1996}, leading to the expectation of a significant EBL component at $\lambda \sim 10-1000\mu$m.  Measurements of this Cosmic Diffuse IR Backround (CDIRB) were notoriously difficult to obtain, owing primarily to the presence of significant zodiacal and galactic foregrounds and the lack of access to this region of the spectrum from the ground. The launch of the {\it Cosmic Background Explorer} \citep[COBE: for an overview, see][]{boggess1992} finally revealed a CDIRB comparable in brightness to the optical EBL, providing a complete census of obscured star formation and AGN activity across cosmic time \citep{hauser1998,kelsall1998,aredt1998,dwek1998,fixsen1998,hauser2001}.

Because the emission mechanisms that generate the CDIRB are intimately connected to galaxy formation and evolution, it provides a powerful observational constraint on models \citep{partridge1967}.  A particularly popular technique for modeling the CDIRB has been backwards evolution \citep[e.g.,][]{rowanrobinson2001,rowanrobinson2009,lagache2003,lagache2004,lagache2005,xu2003,franceschini2008,finke2009b}, in which a parameterized fit to the evolution of the IR luminosity function -- using low-redshift observations as a baseline -- to reproduce a number of observables, including number counts and the CDIRB itself.  A complementary approach has been to use semi-analytic models \citep[SAMs:][]{cole1994,cole2000,somerville1999}, in which the hierarchical growth of structure is tracked by N-body cosmological simulations and baryonic physics -- e.g., star formation, radiative cooling, gas accretion, AGN activity, etc. -- are implemented as simple analytic prescriptions, tuned to match a predetermined set of (usually local) observational constraints.  Mock galaxy catalogs generated by SAMs can then be combined with template spectral energy distributions (SEDs) to construct predictions for number counts at a given wavelength and the CDIRB more generally \citep[e.g.,][]{primack1999,primack2005,primack2008,devriendt2000,baugh2005,swinbank2008}.  Finally, cosmic chemical evolution (CCE) techniques \citep[e.g.,][]{pei1995,pei1999} model the EBL from the optical through IR by solving for the self-consistent evolution of globally averaged quantities -- such as gas depletion and star formation -- in an analogous manner to galactic chemical evolution models \citep[e.g.,][]{tinsley1980}.

Each of these approaches has clear advantages and disadvantages.  Backwards evolution models, while robust and successful in reproducing the observations, must assume simple evolutionary scalings.  Therefore, they contain no physics, and cannot separate out the relative contributions to the CDIRB from different emission mechanisms.  SAMs, by contrast, can determine the relative importance of, e.g. AGN versus star formation, but rely on simple analytic implementations that introduce parameter degeneracies and may not include all of the complexities in the relevant physics -- though some have sought to address this shortcoming by incorporating the results of hydrodynamical simulations of merger driven starbursts and AGN activity into the SAM framework \citep{somerville2008}.  Furthermore, SAMs have also faced difficulty simultaneously matching the observed galaxy mass function and cosmic star formation history \citep{somerville2008}; it is no generally possible to separate these well-known issues from other model elements in driving their predictions for the CDIRB.  Finally, while CCEs do not require a detailed implementation of these complicated baryonic physics, their global nature means they have little to say about specific populations of objects. 

An alternative approach, outlined in detail by \citet{hopkins2007a,hopkins2007b} attempts a semi-empirical
 model of galaxy formation in the context of a merger-driven cosmic cycle: while isolated galaxies dominate the overall cosmic energy budget \citep[however, see discussion in][]{hopkins2010.isolated}, major mergers of gas-rich disks drive an evolutionary sequence through a period of intense star formation and AGN activity, to produce passive elliptical galaxies \citep{sanders1988a,hopkins2006}.  The model connects measurements of the stellar mass function to the dark matter halo population through a halo occupation distribution (HOD) approach \citep[e.g.,][]{peacock2000,scoccimarro2001,berlind2002,berlind2003,kravtsov2004,zehavi2004,zheng2005,zheng2009,brown2008,conroy2009}, and implements baryonic physics by incorporating the results of high-resolution hydrodynamical simulations.  This approach has successfully reproduced the observed quasar luminosity function \citep{hopkins2007a}, the growth of the red sequence \citep{hopkins2007b}, the evolution of massive elliptical galaxies with redshift \citep{hopkins2009.scale.evolve}, bulge-to-disk rations \citep{hopkins2009.gasmorph}, and the IR luminosity function \citep{hopkins2009.ulirg}.  Semi-empirical modeling also represents a fundamental advance over pervious techniques for modeling the CDIRB; it is predictive, incorporates the relevant physics, includes the results of high-resolution hydrodynamical simulations where appropriate, while at the same time is constructed to match the galaxy mass function at all redshifts.  Therefore, we can isolate the physics driving the production of the CDIRB without significant degeneracies model components and parameters. 

In this work, we use this semi-empirical framework to construct a physical model for the origin of the CDIRB.  In addition to providing an important verification of this particular model of galaxy formation more generally, this approach will allow us to unfold the contributions from objects as a function of luminosity and redshift, as well as by emission mechanism -- including star formation in isolated disks, merger-driven starbursts, and AGN activity.  Furthermore, because electron-positron pair production interactions ($\gamma+\gamma \rightarrow e^+ + e^-$) with CDIRB photons is the primary source of attenuation of extragalactic, $\approx$ TeV photons \citep[often referred to as Very High Energy, or VHE $\gamma$-rays;][]{jelly1966,fazio1966,fazio1970,gould1967,stecker1969,stecker1992}, model predictions for the evolution of this background with redshift yields a prediction for the opacity of the Universe to the most energetic photons.  

This work is organized as follows: in \S~\ref{sec:methods} we outline our methodology, in \S~\ref{sec:uncertainty} we summarize the primary sources of uncertainty in the model predictions, in \S~\ref{sec:comparison} we verify that the model is consistent will the relevant observational constraints, in \S~\ref{sec:assumptions} we investigate the result of changing some of the model assumptions, in \S~\ref{sec:interpret} we present predictions for the relative important of different populations of objects in generating the CDIRB, in \S~\ref{sec:gray} we make predictions for the $\gamma$-ray opacity of the Universe, and in \S~\ref{sec:conclusion} we conclude.  Throughout this work, we will make frequent reference to the total IR luminosity which is defined according to convention as the integrated luminosity from 8-1000\micron \citep[e.g.,][]{sanders1996}.  Furthermore, we assume the most recent cosmological parameters from \citet{komatsu2010}: ($\Omega_m,\Omega_\Lambda,h) = (0.26,0.74,0.71)$, the WMAP 7-year mean.  However, it is important to note that varying these within a reasonable range has no noticeable effect to our results.

\section{Methodology}
\label{sec:methods}

In this work, we make extensive use of the semi-empirical model of galaxy evolution presented by \citet{hopkins2007a,hopkins2007b}, and subsequently by \citet{hopkins2009.scale.evolve,hopkins2009.ulirg,hopkins2009.mergers,hopkins2009.gasmorph}; in particular, we utilize the model IRLFs presented by \citet{hopkins2009.ulirg}.  These predictions, which are consistent with the observed galaxy mass function, IRLF, and cosmic star formation history from $z\approx 0-4$, include contributions from the three dominant sources of emission: steady-state star formation, merger-driven starbursts, and obscured AGN activity.  While we refer the reader to the above-mentioned references for details, here were briefly summarize the principal model components and methods and how they are used to construct predictions for the CDIRB.

\subsection{Steady-state star formation}
\label{sec:steadystate}

Given the observed stellar mass function at redshift $z$ \citep[e.g.,][]{bell2003b,fontana2006,arnouts2007,perezgonzalez2008,ilbert2010}, we populate dark matter halos with galaxies following the methodology outlined by \citet{conroy2009}.  These galaxies are then assigned properties consistent with observations of systems at that redshift with comparable stellar masses.  

The gas mass is estimated from observations of the gas mass fractions ($f_g$) of star-forming disks, and is a strong function of mass and redshift \citep[e.g.,][]{bell2001,kannappan2004,mcgaugh2005,shapley2005,daddi2009.fg,tacconi2010}.  It has been shown \citep{hopkins2007a,hopkins2007b,hopkins2009.scale.evolve,hopkins2009.gasmorph} that this evolution is well-described by a simple functional form, motivated by cosmological simulations \citep{keres2005,keres2009}:
\begin{eqnarray}
\nonumber f_g(M_\star\, |\, z=0) &\equiv& f_0 \approx \frac{1}{1+(M_\star/10^{9.15})^{0.4}} \\
f_g(M_\star\, | \, z) &=& f_0 \times \left [1-f_{lb}(z) \left (1-f_0^{3/2} \right ) \right ]^{-2/3}
\end{eqnarray}
where $f_{lb}(z)$ is the fractional look-back time from the present day to $z$ ($f_{lb}\rightarrow 1$ as $z\rightarrow \infty$).  Motivated by the observations listed above, we furthermore assume a constant scatter of $\approx 0.25$ dex independent of stellar mass.  

The size of a given disk galaxy is assigned consistent with the observed size-mass relation \citep{shen2003} and its evolution \citep[e.g.,][]{trujillo2004,ferguson2004,toft2007}.  \citet{somerville2008} found that this too can be represented by a simple analytic fitting form:
\begin{eqnarray}
\nonumber R_e(M_\star\, |\, z=0) &\equiv& R_0 \approx 5.28\, {\rm kpc} \times \left ( \frac{M_\star}{10^{10}} \right )^{0.25} \\
R_e(M_\star \, | \, z) &=& R_0 (1+z)^{-0.6}
\end{eqnarray}
Again, motivated by observations we assume an intrinsic scatter of 0.2 dex.

Finally, given the gas mass and size of a disk galaxy, the rate of steady-state star formation -- i.e., quiescent, or owing primarily to internal gas dynamics rather than external factors \citep[see also][]{hopkins2010.isolated} -- can be estimated using the Kennicutt-Schmidt Law \citep{schmidt1959,kennicutt1998}:
\begin{equation}
\Sigma_{sf} = (1.3\times 10^{-4}) \Sigma_g^{\eta_K}\, M_\odot\, {\rm yr}^{-1}\, {\rm kpc}^{-2} 
\end{equation}
where $\Sigma_g$ is the disk-averaged gas surface density in units of $M_\odot\, {\rm kpc}^{-2}$, $\Sigma_{sf}$ is the disk-averaged star formation rate surface density, $\eta_K = 1.4$ for consistency with observations \citep[see also][]{kennicutt1989,kennicutt1998cal}, and the normalization has been corrected assuming a \citet{chabrier2003} stellar initial mass function (IMF).   Given the assumed IMF, we can convert this star formation rate to the total IR luminosity using the known linear scaling \citep[e.g.,][]{kennicutt1998cal}.

\begin{figure}
\epsfig{figure=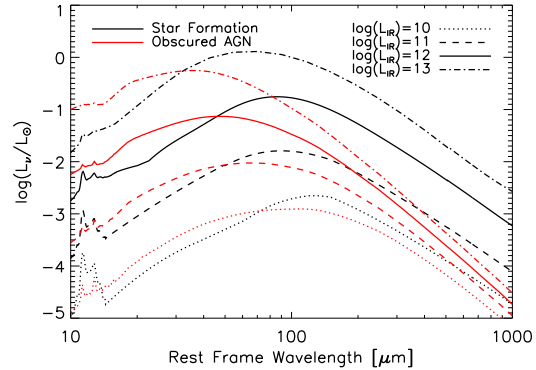,width=76mm}
\caption{Examples of the template SEDs use to connect the model IR luminosity functions to observable quantities.  We show the templates for star-forming systems (black) and AGN (red) at four different luminosities: $\log(L_{IR}/L_{\odot}) = 10$ (dotted line), 11 (dashed line; a LIRG), 12 (solid line; a ULIRG), and 13 (dot-dashed line; a HyLIRG).  The templates for star-forming systems are from the library of \citet{chary2001}, and the AGN are from \citet[][and references therein]{siebenmorgen2004b}}
\label{fig:seds}
\end{figure}

\subsection{Interaction-driven starbursts}
\label{sec:mergers}

In addition to steady-state star formation, close encounters between and mergers of gas-rich galaxies can concentrate large quantities of gas in the nucleus, leading to a dramatic enhancement in the total star formation rate \citep{hernquist1989,barnes1991,mihos1994,mihos1996}.  This owes principally to the effects of the time-varying gravitational potential during close passages: an intrinsic lag in the induced stellar relative to the gaseous bars drains the gas of its angular momentum within a characteristic radius, sending it on a free-fall trajectory to the nucleus \citep{barnes1996,hopkins2009.disksurvival}.  The resulting nuclear gas concentration fuels a massive starburst that can enhance the global starburst by more than an order of magnitude \citep{dimatteo2007,dimatteo2008,cox2008}, and yields characteristic central mass concentrations that have been observed in local elliptical galaxies \citep{hopkins2008.gasrichell,hopkins2009.cores,hopkins2009.cusps}.  

Mergers in our model are tracked via halo-halo merger rates estimated from cosmological N-body simulations \citep[e.g.,][]{stewart2008,fakhouri2008,fakhouri2009}.  We then estimate the galaxy-galaxy merger rate at $z$ as a function of stellar mass,  redshift, and mass ratio ($\mu = M_1/M_2 < 1$) by first correcting for the dynamical friction time delay \citep{boylankolchin2008}, and then convolving with the galaxy population using the HOD model.  Though the exact star formation history of an individual encounter is impossible to determine from our model, the IRLF requires only a knowledge of the distribution of times spent at a given star formation rate, which can be estimated using an ensemble of simulations \citep[e.g.,][]{hopkins2006.sf}.  \citet{hopkins2009.ulirg} found that this can be represented by the fitting function
\begin{equation}
\frac{dt}{d{\rm log}\dot{M}_\star} \sim t_{sb}\,  {\rm exp}\left (-\frac{\dot{M}_\star}{M_{sb}/t_{sb}} \right)
\end{equation}
where $M_{sb}$ is the total mass participating in the starburst, and $t_{sb}\approx 100$ Myr is the characteristic timescale \citep{cox2008}.  It is important to note that tracking the gas content of galaxies is a critical component of the model.  Owing to simple gravitational physics, the burst properties can be represented in simple analytic form (in agreement with the results of hydrodynamical simulations); the total gas mass that is stripped of its angular momentum and participates in the nuclear starburst scales as $M_{sb} \sim (1-f_g) \mu M_g$ \citep{hopkins2009.disksurvival}.  In the context of the model, star formation owing to the interaction-driven burst itself is independent of the gas that retains its angular momentum and continues steady-state star formation in the disk -- a distinction that will become important when characterizing the relative importance of these two modes of star formation (e.g., in \S~\ref{sec:sf}).

\subsection{Obscured AGN}
\label{sec:obscuredagn}

In addition to a nuclear starburst, it has been shown that encounters between gas-rich galaxies will excite AGN activity \citep{dimatteo2005,hopkins2005d,hopkins2005c}.  Furthermore, more recent analyses of hydrodynamical simulations have shown that the peak luminosity of these AGN, corresponding to the Eddington limit of the final black hole mass, is proportional to the binding energy of the bulge in the remnant \citep{hopkins2007theory,younger2008.smbh}.  This bulge mass consists of both the gas mass participating in the burst ($M_{sb}$) and the fraction of the stellar mass which undergoes violent relaxation \citep{lyndenbell1967}.  This fraction is proportional to $\mu M_\star$, again owing to scalings derived from simple gravitational dynamics and in agreement with the results of hydrodynamical simulations  \citep{hopkins2009.disksurvival}.  \citet{hopkins2009.ulirg} find that the relationship between the peak AGN luminosity and these quantities can be conveniently approximated by:
\begin{equation}
L_{peak}^{\rm AGN} \approx 4.6\times 10^{11}\, L_\odot \, \times (1+z)^{0.5} \left (\frac{M_{sb}+\mu M_\star}{10^{10}M_\odot} \right )
\end{equation}
where the redshift evolution reflects the effects of higher gas fractions and smaller scale-lengths in high-redshift disks \citep{hopkins2007theory}.  The distribution of time spent above a given luminosity has been studied extensively in a series of papers \citep[see ][and references therein]{hopkins2006,hopkins2007a}, and is well-described by:
\begin{eqnarray}
\frac{dt}{d{\rm log}L_{bol}} &\approx& 0.22\, Gyr \, \left (\frac{L_{bol}}{L_{peak}^{\rm AGN}} \right )^{\alpha} {\rm exp}\left [ -\left (\frac{L_{bol}}{L_{peak}^{\rm AGN}} \right ) \right ] \\
\alpha &\approx& -0.44+0.21\, {\rm log}\left ( \frac{L_{peak}^{\rm AGN}}{10^{12} L_\odot} \right ).
\end{eqnarray}
Thus, given the merger-rate as a function of redshift, stellar mass, and mass ratio we can estimate the peak luminosity and convolve with the distribution of time above $L_{bol}$ to yield an estimate from the bolometric luminosity function \citep[the whole exercise is presented in][]{hopkins2009.ulirg}.  However, while it is trivial to convert a star formation rate to an IR luminosity, the situation for obscured AGN is rather more complicated.  Therefore, to transform the bolometric AGN luminosity function to the IRLF, we adopt the luminosity-dependent luminosity fraction of \citet{gilli2007} and assume this obscured luminosity is reprocessed into the IR.  Though this is a simplification, we find that it yields similar results to either using the observed column density distribution \citep{ueda2003} or the mean X-ray to IR bolometric correction for obscured AGN \citep{elvis1994,zakamska2004,polletta2006}.  

\begin{figure*}
\begin{center}
\epsfig{figure= 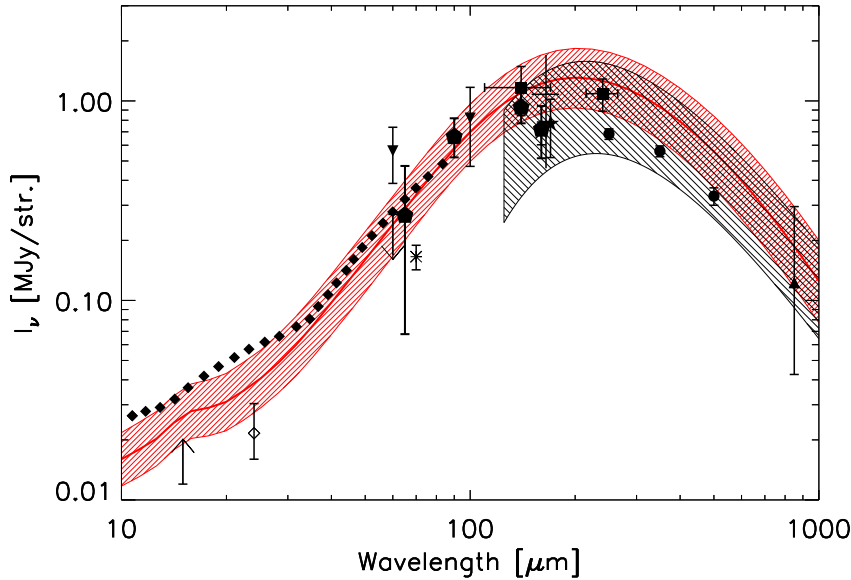,width=120mm}
\end{center}
\caption{The predicted CDIRB from the model (solid red line), including the uncertainty in normalization (red shaded region) as compared to observations: the black shaded region shows the fit and normalization uncertainty from \citet{fixsen1998} to data obtained by the FIRAS experiment on board COBE, the filled diamonds upper limits from $\gamma$-ray absorption \citep[see \S~\ref{sec:gray} for more details;][]{mazin2007}, and and the remaining observations are listed in Table~\ref{tab:obs}}
\label{fig:cirb}
\end{figure*}

\begin{table*}
\label{tab:obs}
\caption{Observational Constraints on the CDIRB}
\begin{center}
\begin{tabular}{cccccccccc}
\hline
\hline
& Wavelength & Symbol & Instrument$^a$ & Reference \\
\hline
\hline
& 1-100\micron & Filled diamond & IACTs & \citet{mazin2007} \\
& 15\micron & Lower limit & ISOPHOT & \citet{chary2001} \\
& 24\micron & Open diamond & MIPS & \citet{papovich2004} \\
& 60\micron & Upper limit & IRAS & \citet{Miville-Deschenes2002} \\
& 60, 100\micron & Inverted triangle & DIRBE & \citet{finkbeiner2000} \\
& 65, 90, 140, 160\micron & Pentagon & AKARI & \citet{matsuura2010} \\
& 70, 160\micron & Asterisk & MIPS & \citet{dole2006} \\
& 100-1000\micron & Black region & FIRAS & \citet{fixsen1998} \\
& 140, 240\micron & Square & DIRBE & \citet{hauser1998} \\
& 150-180\micron & Cross & ISOPHOT & \citet{juvela2009} \\
& 170\micron & Star & ISOPHOT & \citet{lagache2000} \\
& 250, 350, 500\micron & Circle & BLAST & \citet{patanchon2009} \\
& 850\micron & Filled Triangle & SCUBA & \citet{knudsen2008} \\
\hline
\hline
\end{tabular} \\
\end{center}
$^a$ Instruments include: Upper limits from imaging atmostpheric Cherenkov telescopes \citep[IACTs:][]{finley2001,daum1997,hinton2004,cortina2005}, the  Multiband Imaging Photometer for SIRTF \citep[MIPS:][]{rieke2004} on board the Spitzer Space Telescope \citep{werner2004}, the Infrared Astronomy Satellite \citep[IRAS:][]{neugebauer1984}, the  Far-Infrared Absolute Spectrophotometer (FIRAS) and Diffuse Infrared Background Experiment (DIRBE) on board the Cosmic Background Explorer Satellite \citep[COBE: for an overview, see][]{boggess1992}, the Far-Infrared Surveyor \citep[FIS:][]{kawada2007} on board the AKARI Sattellite \citep{murkami2007}, ISOPHOT \citep{Lemke1996} on board the Infrared Space Observatory (ISO), the Balloon-borne Large-Aperture Submillimetre Telescope \citep[BLAST:][]{pascale2008}, and the Submillimetre Common User Bolometer Array \citep[SCUBA:][]{holland1999}.
\end{table*}

\subsection{Predicting the diffuse background emission}
\label{sec:predict}

\begin{figure*}
\begin{center}
\epsfig{figure= 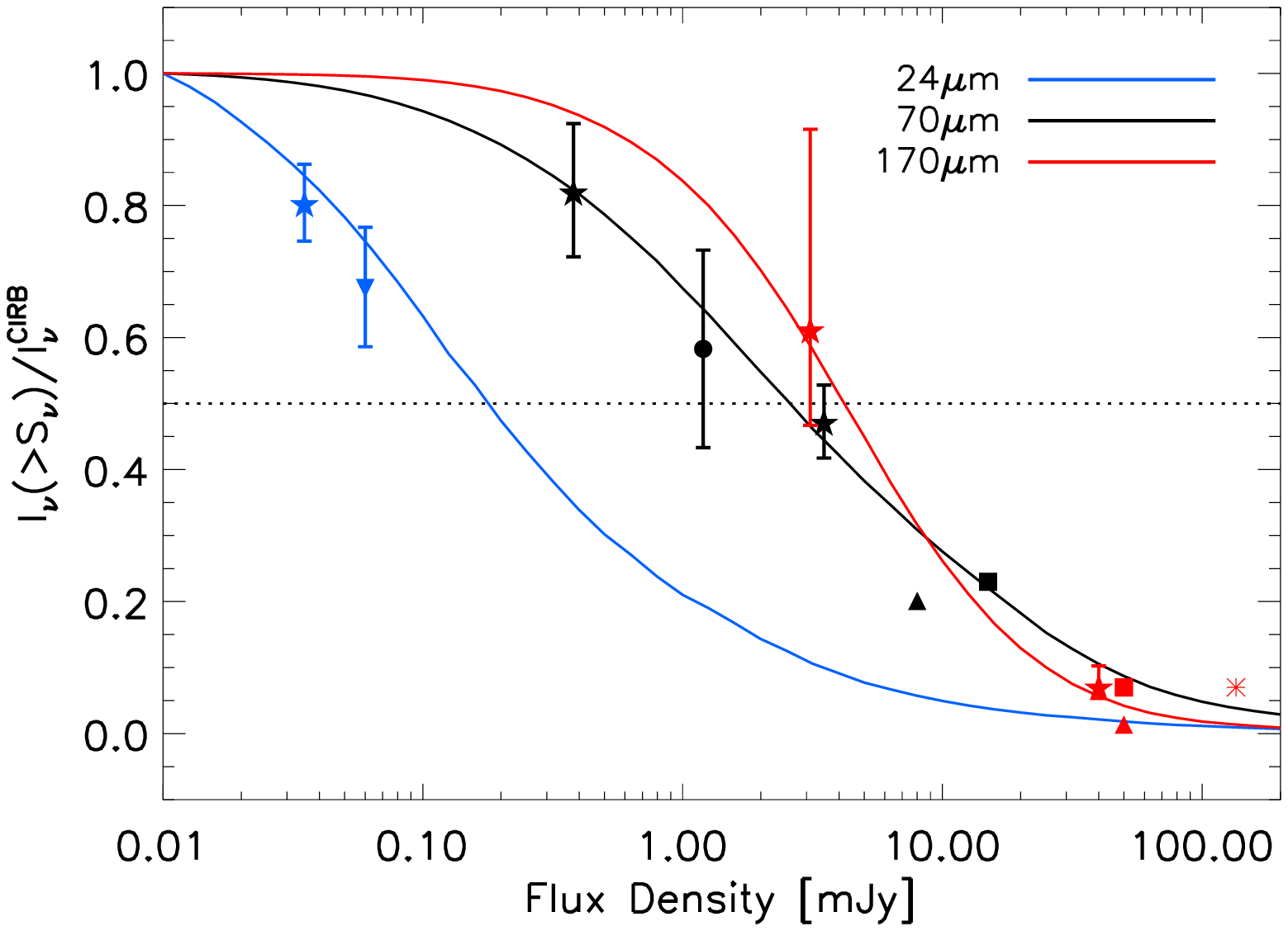,width=76mm}
\epsfig{figure= 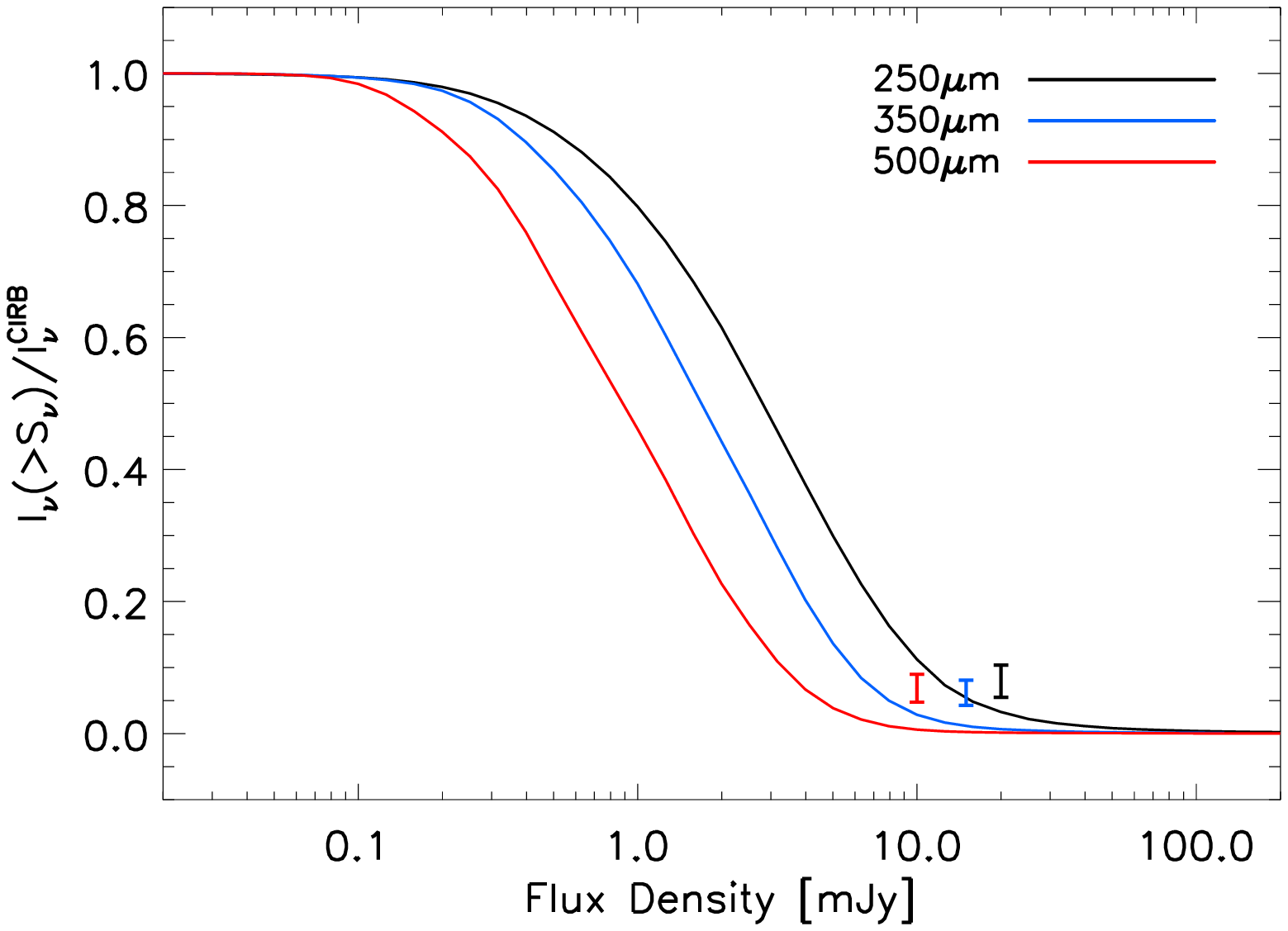,width=76mm}
\epsfig{figure= 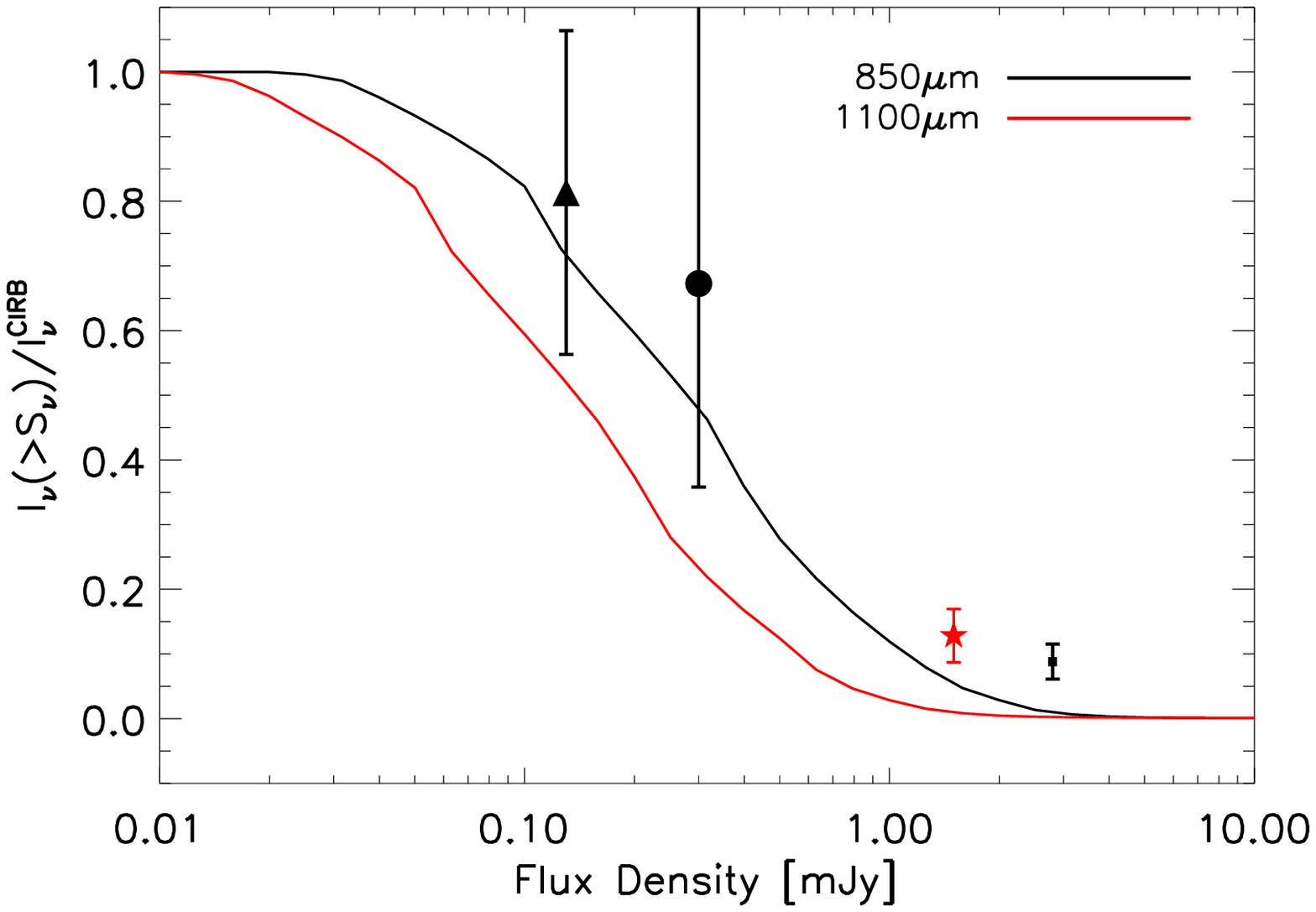,width=76mm}
\end{center}
\caption{The cumulative fraction of the CDIRB resolved as a function of flux density.  The left panel shows model predictions for 24\micron (blue), 70\micron (red), and 160\micron (black) as compared to observations: the black circle is from \citet{frayer2006.goods}, the triangles \citet{frayer2006.fls}, the inverted triangle \citet{dole2006}, the squares \citet{dole2004}, the filled stars \citet{bethermin2010}, and the asterisk \citet{dole2001}.  The right panel shows the BLAST/Herschel bands at 250\micron (black), 350\micron (blue), and 500\micron (red) bands as compared to observations: the vertical lines indicate the error range from \citet{patanchon2009}.  Finally, the bottom panel shows the 850\micron (SCUBA2) and 1100\micron (AzTEC) bands as compared to observations: the square, circle, and triangle are estimates from the number counts of \citet{weiss2009}, \citet{cowie2002}, and \citet{knudsen2008} respectively, and the star is an estimate from the number counts presented by \citet{austermann2010}.}
\label{fig:resolved.cirb}
\end{figure*}

Given a theoretical IRLF  ($dn/d{\rm log}L_{IR}$) and an assumed cosmology, the predicted CDIRB is:
\begin{eqnarray}
\nonumber I_\nu({\rm CDIRB}) = \int dz \int d{\rm log}L_{IR} \, && S_\nu(L_{IR},z) \times \\
&& \frac{dn}{d{\rm log}L_{IR}} \frac{dV}{dzd\Omega}
\end{eqnarray}
where $S_\nu(L_{IR},z)$ is the flux density at frequency $\nu$ in the observed frame for an object with total IR luminosity $L_{IR}$ at redshift $z$, and $dV/dzd\Omega$ is the comoving volume element.  For this work, we derive $S(L_{IR},z)$ for star formation dominated systems (both steady-state and merger-driven star formation) by assigning them templates from the library of \citet{chary2001}, which themselves include models first presented in \citet{dale2001} models.  This template library has been found to reproduce many of the scaling relations observed in local IR-luminous galaxies. For clarity, examples of these templates for three different luminosities are provided in Figure~\ref{fig:seds}.

The choice of SED templates for obscured AGN represents a somewhat more complicated exercise.  While a simple empirical template -- e.g., the local ULIRG Mrk 231, which owes much of its luminosity to a deeply obscured AGN -- can be used, this approach does not account for potential dependence of the effective dust temperature on the luminosity of the system, as is the case for the starburst models \citep{chary2001,dale2001,dale2002,lagache2003}.  Therefore, we choose to utilize a well-studied suite of models for heavily obscured AGN \citep{siebenmorgen1991,siebenmorgen1993,siebenmorgen1992a,siebenmorgen1992b,siebenmorgen2001,siebenmorgen2004a,siebenmorgen2004b}\footnote{Available at {\tt http://www.eso.org/$\sim$rsiebenm/agn\_models/}}, which are parameterized by the outer radius of the dust clouds ($R$), the total emergent IR luminosity ($L_{IR}$), and the visual extinction ($A_V$) -- or alternatively the hydrogen column $N_H/A_V \approx 1.8\times 10^{21}\, {\rm cm}^{-2}$ mag$^{-1}$ \citep{predehl1995}.  In particular, \citet{siebenmorgen2004b} find that optical Type II AGN are best fit by $A_V \approx 16-32\, {\rm mag} \rightarrow \log(N_H/{\rm cm^{-2}}) \approx 22.5-22.8$, and $R\approx 1-2$ kpc.  These columns are also consistent with the observed distribution of column densities in AGN, inferred from X-ray spectroscopy \citep{ueda2003}.  Given these observational constraints, for most of this work we assume $R=1$ kpc and $A_V=16$, or $\log(N_H/{\rm cm^{-2}}) \approx 22.5$; we include examples as a function of luminosity in Figure~\ref{fig:seds}.  

\section{Sources of uncertainty}
\label{sec:uncertainty}

There are three primary sources of uncertainty in the model results: (1) the stellar mass function \citep[also shown by][this dominates the uncertainty in the predicted IRFLs]{hopkins2009.ulirg}, (2) the template SEDs of star formation dominated objects, and (3) the template SEDs of obscured AGN.  The first owes to the dynamic range of observational estimates for the stellar mass function at different redshifts \citep[e.g.,][]{bell2003b,bundy2005,fontana2006,perezgonzalez2008,arnouts2007,ilbert2010} which is the primary model input. As in \citet{hopkins2009.ulirg}, we represent this uncertainty by characterizing the full dynamic range spanned by the observations \citep[see \S~2.1 of][for a detailed description]{hopkins2009.ulirg}, which essentially defines the upper and lower bounds for the model predictions.  This uncertainty naturally increases with luminosity and redshift owing to the lack of large, robust samples of objects which yield a larger effective dispersion among different observational techniques.  In the context of the CDIRB, this yields a $\approx$0.15 dex uncertainty in intensity ($I_\nu$) at $\lambda_{obs} \approx 100$\micron, increasing to $\approx$0.2 dex at longer wavelengths; this owes to larger systematic uncertainties in the IRLF at the high luminosities and redshifts that produce this part of the CDIRB (see \S~\ref{sec:sources}).  It is also important to note that an apparent spike in integral quantities near $z\approx 2$ \citep[e.g., Figs. 4 \& 8 in][]{hopkins2009.ulirg} is an artifact of observational estimates for the stellar mass function at these redshifts and not a physical effect owing to the model itself.

While we find that the observed mass function represents the primary source of uncertainty in model results, for this work we also must consider our lack of knowledge of the template SEDs.  This can be done in a similar manner to the mass function uncertainty, by considering alternative suites of template SEDs and/or considering limiting cases.  For star formation dominated objects -- including both the steady-state and merger-driven modes -- we compare the results of using the template library of \citet{lagache2003} as opposed to our fiducial choice of \citet{chary2001}.  This yields a $\approx$0.05 dex normalization uncertainty independent of wavelength.  We add this in quadrature to our model predictions in Figure~\ref{fig:cirb}, but note that the model uncertainty is still dominated by the dynamic range in stellar mass function estimates.  Obscured AGN again represent a more complicated case, and we consider two methods for determining the systematic uncertainty associated with their contribution to the CDIRB: a limiting case in which the full bolometric luminosity is reprocessed into the IR, and the effects of varying the templated SED model parameters (see \S~\ref{sec:agn}).

\section{Comparison to observations}
\label{sec:comparison}

The primary observational constraint is clearly the observed intensity ($I_\nu$) as a function of wavelength of the CDIRB itself.  In Figure~\ref{fig:cirb} we present the model predictions, including the full range of uncertainty, as compared to a range of observations (see Table~\ref{tab:obs} for references).  We find that the model predictions are consistent with direct observations from the COBE satellite at $\approx 100-1000$\micronend, as well as estimates from number counts in the same wavelength ranges from instruments like IRAS, MIPS, BLAST, and SCUBA.  While many different observations seem to agree in this wavelength range, the mid-IR from $\approx 10-100$\micron has been a source of considerably more controversy owing to the difficulty of effectively subtracting galactic and zodiacal foregrounds and the general lack of access to this wavelength range from the ground. In particular, at 60\micron upper limits from  the power-spectrum of IRAS maps \citep{Miville-Deschenes2002} and the TeV opacity of intergalactic space owing to electron-positron pair production \citep[][and references therein]{stecker1992,dwek2005,mazin2007} are potentially inconsistent with a reanalysis of the DIRBE data by \citet{finkbeiner2000} -- the latest analysis by \citet{matsuura2010} has fairly large systematic uncertainties, and thus does not favor either scenario.  Our model predictions appear not to strongly rule out this measurement, but clearly indicate a preference for a dimmer CDIRB in the mid-IR than they suggest.

Another important test of the model predictions is the relative contributions from individual sources as a function of their flux density.  In Figure~\ref{fig:resolved.cirb} we present this comparison to observations for 24, 70, 160, 250, 350, 500, 850, and 1100\micron data.  It is clear that the model can account for the vast majority of the CDIRB; at 24, 70, 160, and 850\micron we find good agreement with the fraction of emission at those wavelength above a given flux density down to $\approx 10\%$ of the total emission.  At the same time, however, we find that the model cannot account for the $\approx 10-15\%$ of the CDIRB owing to the brightest sources at $\lambda_{obs} \gsim 250$\micronend.  This is a well-known problem, exemplified by the difficulty semi-analytic models have found in reproducing the bright 850\micron counts \citep{baugh2005,swinbank2008}.  There are a number of potential solutions have been proposed in the literature, including cosmological gas accretion \citep{dave2009} and a more sophisticated calculation of the IR SED including full radiative transfer \citep[e.g.,][]{narayanan2009.smg} combined with the well-known effects of cosmic variance in these rather small-area surveys \citep[see discussion in e.g.,][]{weiss2009,austermann2010}.  However, because the model agrees with observations over $\gsim 90\%$ of the CDIRB emission, we do not believe this discrepancy materially affects our results.  Modeling the bright end of the luminosity function and the number counts more generally is, however, clearly a very interesting problem in its own right and a full analysis is currently ongoing (C. C. Hayward et al., in preparation).  

\section{Changing the model assumptions}
\label{sec:assumptions}

That our model is entirely consistent with available observational constraints is very encouraging.  However it is important to examine how its predictions for the shape and normalization of the CDIRB are affected by changing our assumptions.  For example, strong evolution in the gas content of galaxies and the constancy of the Kennicutt-Schmidt Law with redshift -- both critical components of the model -- may themselves drive the model predictions.  Relaxing them provides insight into the origin of the CDIRB, as well as a test of the robustness of our results.  It is also important to note that while we could just as well vary the model assumptions around the sizes and evolution of disks, the trend of obscured fraction with AGN luminosity and redshift, and evolution of the IMF with redshift, \citet{hopkins2009.ulirg} found that these were not a significant source of uncertainty in their predictions.

\begin{figure}
\epsfig{figure= 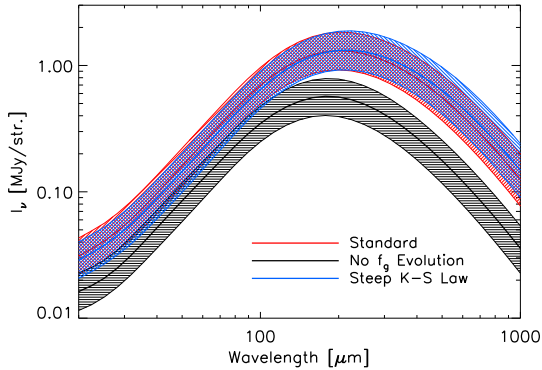,width=76mm}
\caption{The predicted CDIRB under three different sets of model assumptions: the fiducial model (red), a steeper Kennicutt-Schmidt Law ($\eta_K \approx 1.6$ rather than 1.4; blue), and no significant evolution in the gas content of disk galaxies with redshift (black).}
\label{fig:assumptions}
\end{figure}

In Figure~\ref{fig:assumptions} we present the model predictions for the SED of the CDIRB for two scenarios: one in which the gas content of disks does not evolve with redshift, and another in which we assume a somewhat steeper Kennicutt-Schmidt Law \citep[the effects of both of these variations on the IRLFs are presented by][]{hopkins2009.ulirg}.  Clearly the most dramatic differences arise from ignoring the redshift evolution of gas fractions at fixed stellar mass.  This systematically suppresses the Universal star formation rate at intermediate to high redshift relative to the fiducial model by lowering the gas mass of those disks.  Its effect on the predicted CDIRB is two-fold: first the normalization -- particularly at $\gsim 100\mu$m -- is substantially lower, reflecting the systematic decrease in Universal star formation rate density, and therefore IR luminosity density at $z>0$ relative to the fiducial model; second, it shifts the peak of emission to shorter wavelength, owing to the preferential suppression of high-redshift star formation which is produced by emission longward of the peak in a typical starburst SED \citep[$\lambda_{obs} \approx 60-100\mu$m;][]{sanders1996}.  While there is strong observational evidence that high-redshift disks are preferentially gas-rich \citep[e.g.,][]{bell2001,kannappan2004,mcgaugh2005,shapley2005,daddi2009.fg,tacconi2010}, this exercise demonstrates our model requires a systematic and substantial evolution in their gas content to remain consistent with observations.  Put another way, we can interpret the shape and normalization of the CDIRB as reflecting this increase with redshift of the typical gas fraction of galaxies at constant stellar mass.

\begin{figure}
\epsfig{figure= 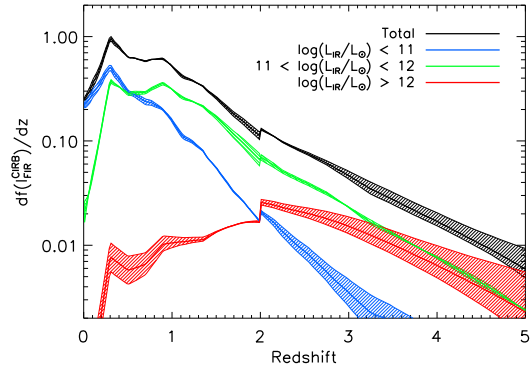,width=76mm}
\caption{The differential contribution to the predicted CDIRB as a function of redshift including all objects (black) and separated into three luminosity bins: `normal' (blue), LIRGs (green), and ULIRGs (red).  It is also important note that the apparent discontinuity near $z\approx 2$ is an artifact of the observational estimates for the stellar mass function at those redshifts adopted for the model, rather than a physical effect.}
\label{fig:highz1}
\end{figure}

Though local star-forming galaxies are well-fit by a Kennicutt-Schmidt index of $\eta_K \approx 1.4$, there is some observational evidence for a steeper relation at high-redshift and in more extreme environments \citep{bouche2007,bothwell2010}.  Therefore, we consider a model which is otherwise identical to the fiducial set of assumptions, but with $\eta_K \approx 1.6$ (renormalizing the relation to avoid overproducing the local star formation rate).  \citet{hopkins2009.ulirg} found that the primary effect of this change was somewhat more star formation in the most massive systems at high-redshift.  The result is a set of IRLFs that are largely unchanged considering the systematic uncertainties in the model, though with a slightly higher contributions form the high-luminosity, high-redshift tail.  In the context of the CDIRB, the results are again similar with a bit more emission at $\lambda_{obs} \gsim 300\mu$m -- a region of the SED which we will see in the proceeding section has the most significant contribution from high-redshift ULIRGs.  But again, considering the systematic uncertainties, we find that the model predictions are relatively insensitive to this change in $\eta_K$.

\section{Interpreting the cosmic diffuse infrared background}
\label{sec:interpret}

\subsection{Sources of emission}
\label{sec:sources}

Given that our model predicts a CDIRB that is consistent with the observational constraints, we can use it to interpret the origin of the emission.  In particular, we can separate out the relative contribution from sources at different redshifts and in three different luminosity bins: `normal' systems ($L_{IR} < 10^{11}$ $L_\odot$), luminous infrared galaxies (LIRGS: $10^{11}\, L_\odot \leq L_{IR} < 10^{12}\, L_\odot$), and ultralumionous infrared galaxies (ULIRGs: $L_{IR} > 10^{12}$ $L_\odot$).  In Figure~\ref{fig:highz1} we show the differential contribution of each to the integrated background emission as a function of redshift.  The model shows that observed intensity of the CDIRB is dominated by sources at $z\sim 0.5-1$ and consists primarily of `normal' systems and LIRGs.  In general, since $dI_{IR}/dz \sim \dot{\rho}_{\ast}(z)(dV/dz)/D_L^2(z)$ -- where $\rho_\star$ is the cosmic star formation rate density, $dV/dz$ is the comoving volume element, and $D_L$ is the luminosity distance -- this is consistent what one would expect from observations of the IR luminosity density and cosmic star formation history, which have been found to peak at $z\sim 1$ \citep[][and references therein]{ahopkins2006}; at those redshifts the IR luminosity density is produced by roughly equal contributions from `normal' systems and LIRGs \citep{lefloch2005,magnelli2009}.  The model furthermore predicts that ULIRGs will become more important than `normal' systems in producing the CDIRB only at $z\gsim 2$, and will equal that of LIRGs at even higher redshift $z\gsim 3$.  However, the vast majority of the total CDIRB emission comes from objects a low to moderate redshifts, with $\approx 50\%$ arising from contributions at $z\lsim 1$ and 90\% from $z\lsim 2$.  Again, the discontinuities in these quantities at $z\approx 2$ are artifacts of the observed galaxy mass functions adopted, rather than a physical effect. 

\begin{figure}
\epsfig{figure= 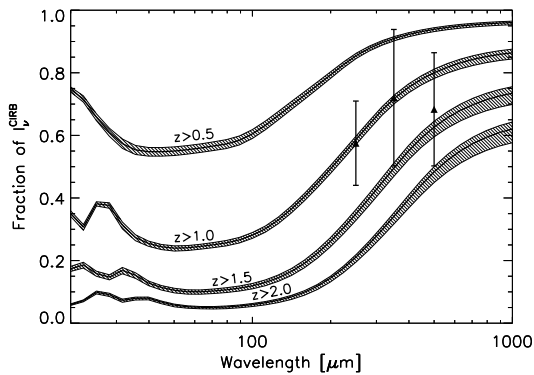,width=76mm}
\caption{Model predictions for the fraction of CDIRB flux density as a function of wavelength owing to sources at (top to bottom) redshift of $z > 0.5$, 1.0, 1.5, and 2.0 including objects of all luminosities.  For comparison we show BLAST observations of the fraction of CDIRB light contributed by sources at $z\gsim 1$ estimated via a stacking analysis \citep{marsden2009}.}
\label{fig:highz2}
\end{figure}

\begin{figure}
\epsfig{figure= 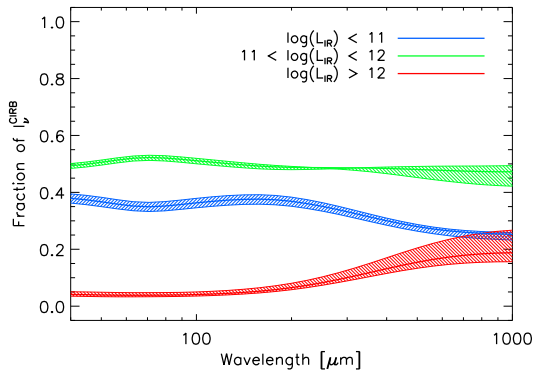,width=76mm}
\caption{Model predictions for the fraction of CDIRB flux density as a function of wavelength owing to sources at in three different luminosity bins: `normal systems' (blue), LIRGS (green), and ULRIGs (red).}
\label{fig:highz3}
\end{figure}

In Figure~\ref{fig:highz2} we isolate the contribution from sources above a given redshift threshold.  This can then be compared to observational estimates measured via a stacking analysis using BLAST data at 250, 350, and 500\micron which found roughly half of the CDIRB at those wavelengths was contributed by sources at $z\gsim 1$ \citep{marsden2009,devlin2009}.  We also find a greater contribution from high-redshift systems at longer wavelengths past the peak of the CDIRB ($\lambda \gsim 200$\micronend); this again makes sense considering the peak in emission for IR-luminous systems that dominate the local IR luminosity density is $\approx 60-120$\micron \citep[see review by][]{sanders1996}, and therefore longer wavelength emission in the CDIRB must arise primarily from redshifting this peak to longer wavelength.  

We perform a similar exercise in Figure~\ref{fig:highz3}, breaking down the CDIRB into contributions from the three luminosity bins.  We find that LIRGS contribute $\approx 50\%$ of the observed emission at all wavelengths; `normal' systems contribute $\approx 40\%$ at $\lambda_{obs} \approx 20\mu$m decreasing to $\approx 30\%$ at $\lambda_{obs} \approx 1000\mu$m while the contribution from ULIRGs increases from $\approx 5\%$ to $\approx 20\%$ over the same interval.  This owes to the strong evolution in the IRLF of ULRGs, yielding a much more important contribution at $z\gsim 1$ than at low-redshift \citep{lefloch2005,magnelli2009} and thus contributing a much larger fraction of the CDIRB at wavelengths that are dominated by emission from high-redshift objects.

\subsection{Steady-state versus interaction-driven star formation}
\label{sec:sf}

\begin{figure}
\epsfig{figure= 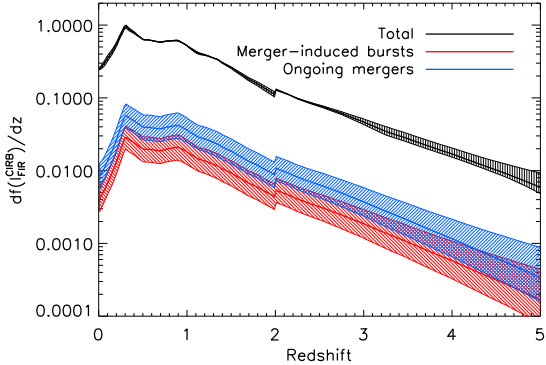,width=76mm}
\caption{The differential contribution from merger-induced bursts (red) relative to the total (black) CDIRB as a function of redshift.  It is important to note that in this case merger-driven refers to star formation directly attributable to the merger event, rather than all star formation in ongoing merger.  The total contribution from ongoing mergers, including the steady-state mode is also shown (blue).}
\label{fig:sf1}
\end{figure}

\begin{figure}
\epsfig{figure= 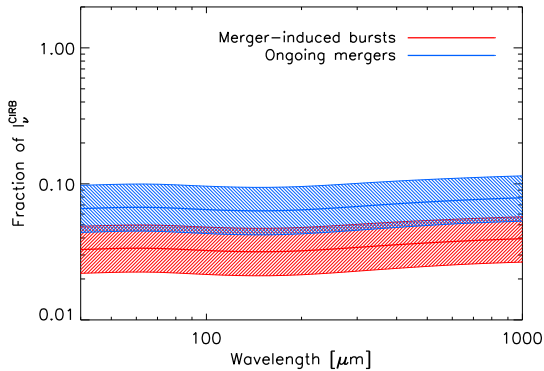,width=76mm}
\caption{Same as Figure~\ref{fig:sf1}, but for the fractional contribution function of the observed CDIRB flux density as a function of wavelength.}
\label{fig:sf2}
\end{figure}

The primary emission mechanism for producing the CDIRB is star formation, in which dust associated with periods and regions of active star formation reprocesses their UV-optical light into thermal emission.  The model predictions presented in this work allow us to unfold the relative importance of two different modes of star formation: steady-state and interaction-driven (see \S~\ref{sec:steadystate} and \ref{sec:mergers} for an overview).  In Figure~\ref{fig:sf1}, we show the differential fraction of the total CDIRB emission contributed as a function of redshift, including both the total and that arising from merger-driven bursts.  We find that mergers generate only a small fraction, on the order of a few percent, of the emission at all redshifts.  Steady-state star formation therefore contributes the vast majority of the CDIRB flux density observed today.  The same is true at all wavelengths (see Figure~\ref{fig:sf2}), with merger-driven star formation accounting from only $\approx 2-5\%$ of the total emission.

This owes trivially to the fact that, in our model merger-driven bursts account for a similarly small fraction of the IR luminosity density, which is consistent with a number of observational constraints \citep[out to $z\sim 1$;][]{brinchmann1998,bell2005,lotz2008,jogee2009,robaina2009,lopezsanjuan2010} and semi-empirical estimates \citep{hopkins2010.sfh}.  It is important to note, however, that in this context 'merger-driven' refers only to the star formation owing directly to the interaction beyond what the two disks would have formed in isolation.  When all on the star formation, including the steady-state mode, is taken into account, systems that would be identified as ongoing mergers \citep[e.g., via their morphology;][]{lotz2008b,lotz2009a} can be a factor of 1.5-3 higher (also shown in Figures~\ref{fig:sf1} and \ref{fig:sf2}).  However, even in this limiting cases mergers will never contribute more than $\approx 10\%$ of the CDIRB emission.

\subsection{The role of obscured AGN}
\label{sec:agn}

Though star formation clearly dominates the CDIRB, observational constraints on the importance of obscured AGN remain a topic of some controversy and debate.  For example, a number of authors \citep{almaini1999,risaliti2000.cirb,blain2002.cirb,barger2001} have claimed that AGN contribute of order 5-10\% of the emission, making them comparable in importance to merger-driven starbursts while others \citep{severgnini2000,silva2004,fardal2007} claim they make a trivial contribution.  Again, our model predictions can be used to separate out obscured AGN, and estimate their relative importance in producing the CDIRB.

\begin{figure}
\epsfig{figure= 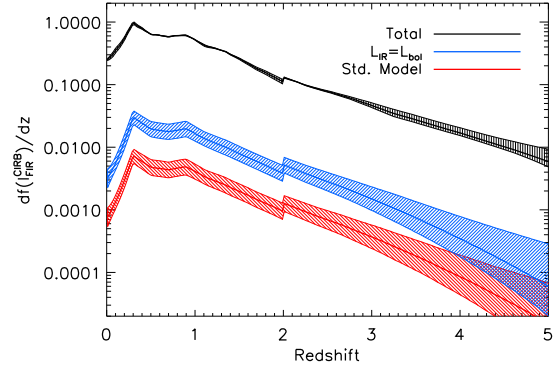,width=76mm}
\caption{The differential contribution to the CDIRB from obscured AGN relative to the total (black) for two different cases: a limiting case in which the total bolometric luminosity of all AGN is assumed to be reprocessed into the IR (blue), and the prediction from the fiducial model assuming the template SED library and model parameters that best-describe Type II AGN \citet{siebenmorgen2004b}.}
\label{fig:agn1}
\end{figure}

In Figure~\ref{fig:agn1} we show the differential fraction of the total CDIRB as a function of redshift for the fiducial model (red shaded region) as well as a limiting case in which we assume that 100\% of the AGN's bolometric luminosity is reprocessed into the IR at $10\mu{\rm m}\lsim \lambda \lsim 100\mu{\rm m}$.  Therefore, our fiducial model predicts that AGN contribute $0.7\pm 0.2\%$ of the total CDIRB emission, with a firm maximal contribution of $3\pm 1\%$ in the limiting case.  This results from a simple energetic argument: AGN do not produce enough photons to substantially affect the CDIRB.

\begin{figure*}
\begin{center}
\epsfig{figure= 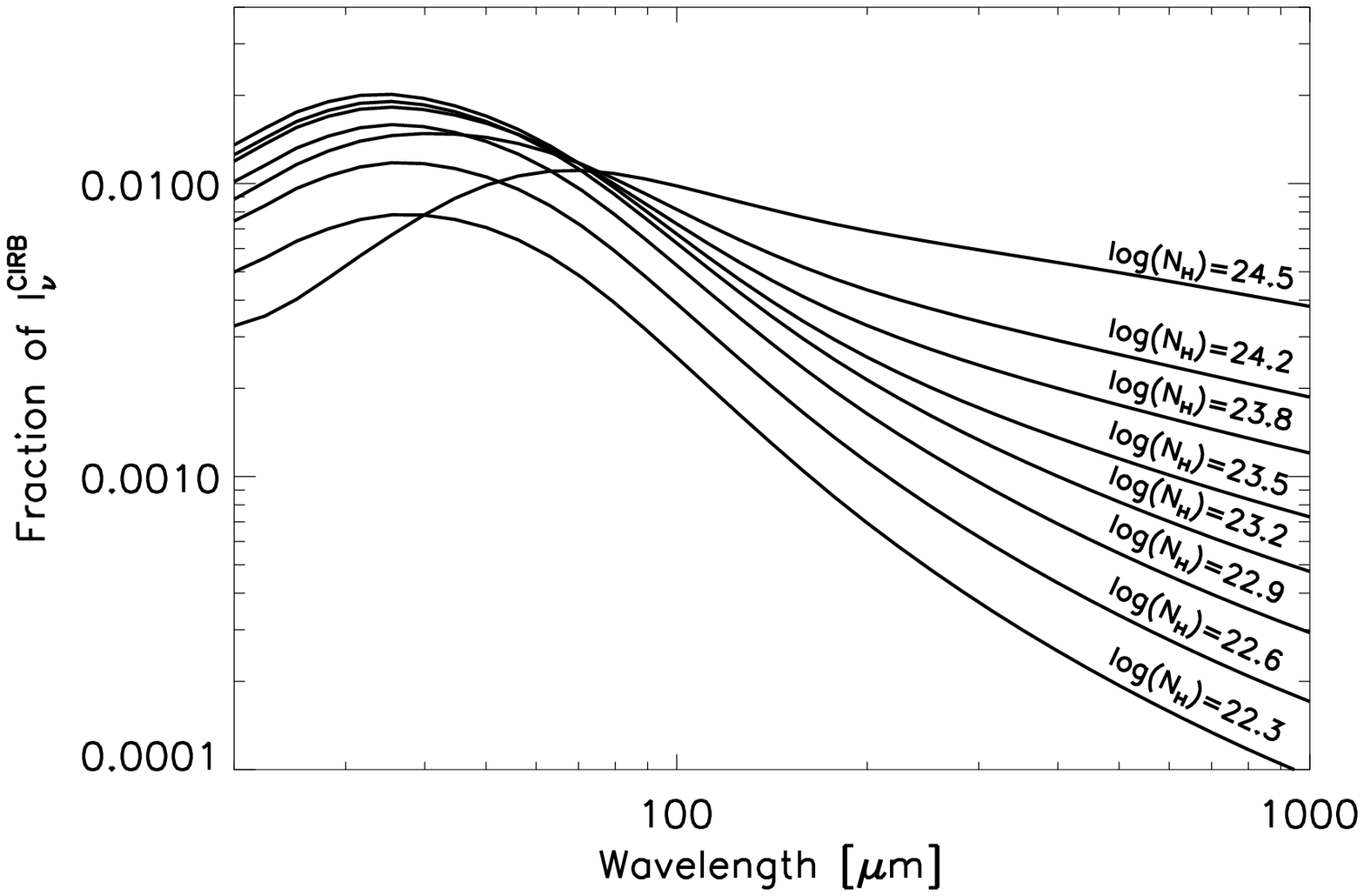,width=76mm}
\epsfig{figure= 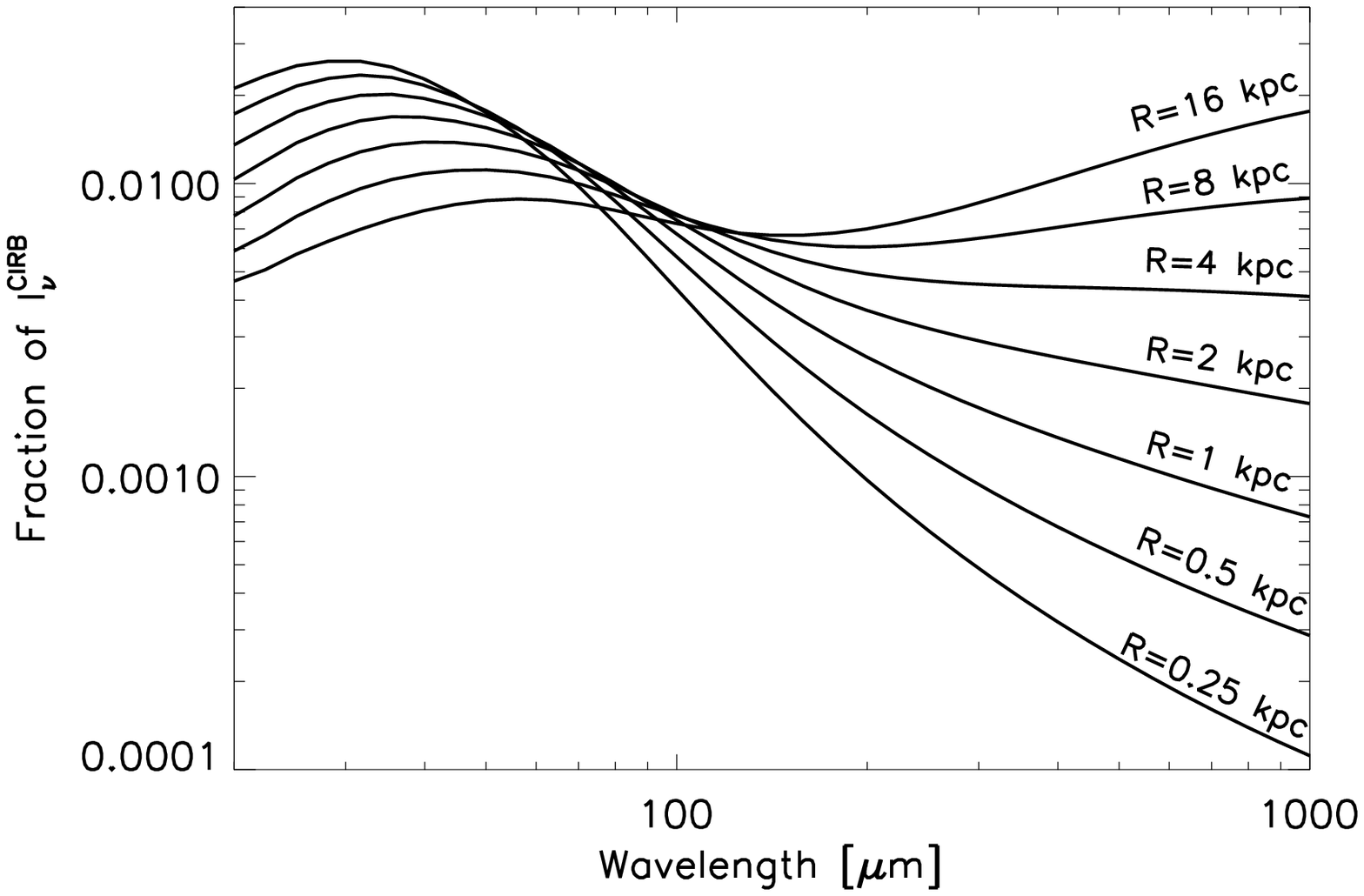,width=76mm}
\end{center}
\caption{The fraction of the CDIRB flux density as a function of wavelength owing to obscured AGN for the full range of SED model parameters.  The left panel shows the contribution at fixed $R=1$ kpc for (bottom to top) $A_V=1,\, 2,\, 4,\,\ldots 128$ ( $\log(N_H/{\rm cm^{-2}}) = 22.3,\, 22.6,\, 22.9,\ldots 23.5$).  The right panel shows the contribution at fixed $A_V=16$ ($\log(N_H/{\rm cm^{-2}})=22.5$) for $R=0.25,\, 0.5,\, 1.0,\, 16$ kpc.}
\label{fig:agn2}
\end{figure*}

Though the model predicts the AGN contributes $\lsim 1\%$ of the total CDIRB emission, it is in principal possible that it could contribute relatively more at certain wavelengths.  Therefore, in Figure~\ref{fig:agn2} we present the fraction contribution from obscured AGN to the CDIRB as a function of wavelength.  In this case, rather than examining the bolometric limiting case as we did previously (which by definition ignores wavelength dependence) we can instead varying the model parameters about our fiducial choice ($A_V=16$ and $R=1$ kpc) to examine their effect on the prediction.  In the left panel, we hold $R$ constant while varying $A_V$, and in the right panel we hold $A_V$ constant while varying $R$.  In most cases, the AGN contribution will peak at $\sim 30-40\mu$m owing to its relatively warmer IR SED as compared to star-forming systems of the same IR luminosity (see examples in Figure~\ref{fig:seds}).  Only in the most extreme column densities and largest sizes will the template SED be cold enough to contribute more at longer wavelengths.  However, while the AGN can contribute somewhat more than $\sim 2\%$ at the shortest wavelengths, they are clearly subdominant in general, even under the most extreme assumptions of obscuration and size.

\section{The $\gamma$-ray opacity of the Universe}
\label{sec:gray}

The CDIRB and diffuse backgrounds more generally also represent the primary source of opacity to very high energy $\gamma$-rays in intergalactic space \citep{jelly1966,fazio1966,fazio1970,gould1967,stecker1969,stecker1992}. When the mass-energy of photons is comparable to the rest-mass of electrons, they can interact via electron-positron pair production: $\gamma+\gamma \rightarrow e^+ + e^-$.  The cross section for interactions between a photon with energy $E_\gamma$ with a background photon at energy $\epsilon = h \nu$ at redshift $z$ is:
\begin{eqnarray}
\nonumber \sigma_{\gamma\gamma}(E_\gamma,\epsilon,\mu&)& = \frac{3\sigma_T}{16} (1-\beta^2) \times \\
&& \left [2\beta(\beta^2-2)+(3-\beta^4){\rm ln}\left (\frac{1+\beta}{1-\beta} \right ) \right ] \\ 
\beta &\equiv& \sqrt{1-\frac{\epsilon_{th}}{\epsilon}} \\
\epsilon_{th} &=& \frac{2(m_e c^2)^2}{E_\gamma (1-\mu)}.
\end{eqnarray}
where $\sigma_T$ is the Thompson cross section, $\epsilon_{th}$ is the threshold energy, $\mu \equiv \cos{\theta}$ where $\theta$ is the angle of incidence, and $n_\epsilon \equiv dn_\gamma/d\epsilon$ is the differential comoving number density of background photons at $z$ given by:
\begin{equation}
\epsilon^2 n_\epsilon(\epsilon,z) = \frac{4 \pi}{c} \nu I_\nu(\nu,z).
\end{equation}

To calculate the optical depth of a photon with energy $E_\gamma$ in the observers frame that was emitted by a source at redshift $z_{em}$, simply integrates along the line of sight:
\begin{eqnarray}
\nonumber && \tau_{\gamma\gamma}(E_\gamma,z_{em}) = \\ \nonumber \\
\nonumber && \int_{0}^{z_e} dz' c \frac{dt}{dz}  \int_{-1}^{+1}d\mu' \frac{1-\mu'}{2} \int_{\epsilon_{th}}^{\infty} d\epsilon' n_\epsilon(\epsilon',z') \sigma_{\gamma\gamma}(E_\gamma,\epsilon',\mu') \\
\end{eqnarray}
where $c dt/dz$ is the comoving line-of-sight distance \citep[see][]{hogg1999}.  Since the primary emission mechanism at these energies is thought to be synchrotron self-Compton, and therefore is well-described by a simple power-law \citep[e.g.,][]{aharonian2002}, one can used $\sigma_{\gamma\gamma}$ inferred from the observed GeV-TeV SEDs of $\gamma$-ray sources to constrain the intensity of the diffuse background \citep{stecker1992,stecker1993,stecker2007,dwek1994,dwek2005,dejager1994,madau1996b,funk1998,primack1999,bullock1999,renault2001,aharonian2002,aharonian2006,schroedter2005,katarzynski2006,mazin2007,georganopoulos2008,finke2009b}. In general, for a uniform background of photons, the cross section peaks at $\beta=0.7$ which corresponds to a background photon with wavelength $E_\gamma \epsilon \approx 4 (m_e c^2)^2$ or $(\lambda_\epsilon/{\rm \mu m}) \approx 1.2\times (E_\gamma/{\rm TeV})$.  The diffuse background has a trough in the near-IR before rising steeply out to $\lambda_{obs} \approx 200\mu$m \citep{hauser2001,lagache2005}.  Therefore, the the opacity of the Universe to $\sim$few$\times$TeV photons (typically referred to Very High Energy, or VHE $\gamma$-rays) is generically dominated by the mid-IR component of the diffuse background.  

\begin{figure}
\epsfig{figure= 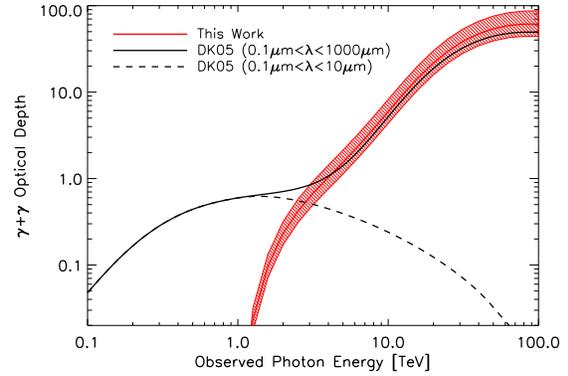,width=76mm}
\caption{The opacity of the local Universe to electron-positron pair production interactions ($\gamma+\gamma \rightarrow e^+ + e^-$) as a function of the observed energy ($E_\gamma$), assuming a source at $z=0.03$.  The results assuming the mean CDIRB from \citet{dwek2005} are shown as a solid black line, and the same only including the optical through near-IR component of the background photons ($\lambda \lsim 10\mu$m) as a dashed line.  We also show the model predictions (red shaded region), which includes background photons with $\lambda \gsim 10\mu$m.  Over the range where the CDIRB dominates the $\gamma+\gamma$ opacity ($E_\gamma \gsim 1$ TeV), the model provides a good match to the \citet{dwek2005} results.}
\label{fig:tgg1}
\end{figure}

\begin{figure}
\epsfig{figure= 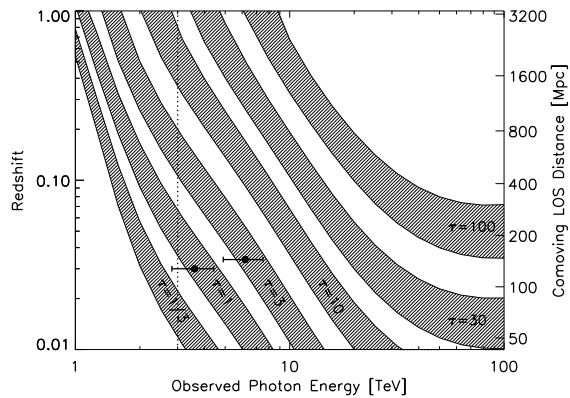,width=76mm}
\caption{Model predictions for the absorption edges due to electron-positron pair production as a function of redshift or line-of-sight comoving distance (left/right absciss) and observed photon energy ($E_\gamma$) for a range of optical depths: $\tau_{\gamma\gamma} = \frac{1}{3}$, 1, 3, 10, 30, and 100.  We also include the cut-off energies observed for two local TeV sources \citep{aharonian1999,aharonian2002a}.}
\label{fig:tgg2}
\end{figure}

Since our model predicts the CDIRB intensity, and therefore number density, we can invert this procedure and instead provide predictions for opacity of the Universe due to pair creation interactions as a function of $E_\gamma$ and $z_{em}$ \citep[see also][]{primack2005,primack2008,franceschini2008,finke2009}.   In Figure~\ref{fig:tgg1}, we compare the model predictions to estimates of $\tau_{\gamma\gamma}$ using the 'average' local CDIRB model of \citet{dwek2005}.  We find good agreement between the two, and also confirm that by only including the IR component of the diffuse background emission, we are essentially capturing the opacity of the Universe to observed photons with $E_\gamma \gsim 2-3$ TeV; below these energies the model predictions should be considered a lower limit.

Figure~\ref{fig:tgg2} shows the absorption edges predicted by the model for a number of optical depths ranging from $\tau_{\gamma \gamma} = 1/3-100$.  Two TeV sources have been observed to have exponential cut-offs in the VHE $\gamma$-ray SEDs \citep{aharonian1999,aharonian2002a} which are consistent with the model predictions for $\tau_{\gamma\gamma} \gsim 1$ at those redshifts.  These results suggest that the Universe is very optically thick to TeV photons beyond a redshift of $z_{em} \approx 0.2$ with $\tau_{\gamma\gamma} \gsim 3$ for $E_\gamma \gsim 3$ TeV and $\tau_{\gamma\gamma} \gsim 10$ for $E_\gamma \gsim 8$ TeV.  Furthermore, at $z_{em} \gsim 0.5$ the Universe is optically thick down to $\approx 1$ TeV, and $\tau_{\gamma\gamma} \approx 10$ down to $\approx$ 3 TeV.  At higher energies ($E_\gamma \gsim 10$ TeV) the Universe is very optically thick ($\tau_{\gamma\gamma}\approx 3$ all the way down to $z_{em} \approx 0.01$.  Though significant improvements in sensitivity, particularly at $\gsim 10$ TeV, are planned in the next generation of ICATs \citep[e.g., the Cherenkov Telescope Array;][]{hermann2008}, these results suggest that the high optical depth to pair creation might pose a significant problem for VHE observations at even somewhat low redshifts.  One can, however, expect to observed $\approx $ TeV photons out to $\approx 100-150$ Mpc.

\section{Conclusions}
\label{sec:conclusion}

We present a physical model for the origin of the CDIRB utilizing the semi-empirical framework of \citet{hopkins2009.ulirg}.  The model tracks three distinct sources of IR emission: steady-state star formation, interaction-induced starbursts, and obscured AGN activity.  We also include all the relevant systematic uncertainties, which are dominated by the dynamic range in observational estimates of the stellar mass function (and increasing with redshift).  The IRLFs generated by this model, combined with a library of template SEDs for starbursts \citep{dale2001,chary2001,lagache2003} and obscured AGN \citep{siebenmorgen1991,siebenmorgen1993,siebenmorgen1992a,siebenmorgen1992b,siebenmorgen2001,siebenmorgen2004a,siebenmorgen2004b}, provide an excellent match to observations of the CDIRB from $\lambda_{obs} \approx 10-1000\mu$m.  In contrast of alternative techniques -- including backwards evolution \citep[e.g.,][]{rowanrobinson2001,rowanrobinson2009,lagache2003,lagache2004,lagache2005,xu2003,franceschini2008,finke2009b}, semi-analytic \citep{primack1999,primack2005,primack2008,devriendt2000}, and CCE \citep{pei1995,pei1999} models -- our approach provides a robust context in which to examine the physics its production.

By varying the model assumptions, we find that the normalization and peak of the CDIRB emission are determined in large part by the strong redshift evolution of the gas content of steady-state star-forming galaxies.  Without this evolution, the model vastly underpredicts the total intensity of CDIRB and peaks at shorter wavelengths than is observed, owing to the lack of a significant contribution form high-luminosity and high-redshift disks.  Therefore, the observed CDIRB can be taken to reflect this strong evolution in the gas content of disks -- an effect which has been seen in numerous samples of individual objects \citep[e.g.,][]{bell2001,kannappan2004,mcgaugh2005,shapley2005,daddi2009.fg,tacconi2010}.

We can also use the model predictions to determine the relative importance of IR-luminous galaxies in producing the CDIRB as a function of redshift, luminosity, and emission mechanism.  The model results indicate the following:
\begin{enumerate}
\item The CDIRB is primarily produced by equal contributions form objects at $z\sim 0.5-1$ and $z\gsim 1$, in agreement with recent observations by the BLAST experiment.  However, in the observer's frame wavelengths past the peak contain an larger contribution from $z\gsim 1$ than those at shorter wavelengths.   
\item Most of the CDIRB is contributed by normal galaxies and LIRGs, though the ULIRG contribution becomes significant at high-redshift $z\gsim 2$.  This also leads to a larger contribution from ULIRGs at wavelengths past the peak.
\item Ongoing mergers contribute less than 10\% of the total emission at all wavelengths, with less than $\sim 1-3\%$ produced by the merger-driven burst itself.
\item Obscured AGN account for $\lsim 1\%$ of the CDIRB at all wavelengths, with a strong upper limit at 4\% of the total emission.
\end{enumerate}
In the future, these predictions can be tested via deep, wide surveys with the {\it Herschel Space Telescope} (e.g., the Herschel Multi-tiered Extragalactic Survey, or HerMES)\footnote{{\tt http://astronomy.sussex.ac.uk/$\sim$sjo/Hermes/}} combined with stacking analyses similar to those presented by \citet{devlin2009} and \citet{marsden2009}. 

The CDIRB also represents the primary source of opacity for VHE $\gamma$-rays, owing to electron-positron pair production interactions \citep[$\gamma+\gamma \rightarrow e^+ + e^-$;][]{jelly1966,fazio1966,fazio1970,gould1967,stecker1969,stecker1992}.  Thus, our results also provide predictions for the opacity of the Universe to the most energetic photons.  We find that the model predictions are consistent with high-energy cutoffs for TeV sources at $z\sim 0.03$ \citep{aharonian1999,aharonian2002a}.  They also indicate that while the Universe is highly opaque to $E_\gamma \gsim 10$ TeV photons ($\tau_{\gamma\gamma} \gsim 10$ at $z\approx 0.06$), with the next generation of ICATs we can reasonably expect to detect sources out to $\sim 50-150$ Mpc.

\subsection*{Acknowledgements}

Thanks to Michael Kuhlen, Desika Narayanan, and Chris Hayward for helpful converstaions.  JDY acknowledges support from NASA through Hubble Fellowship grant \#HF-51266.01 awarded by the Space Telescope Science Institute,which is operated by the Association of Universities for Research in Astronomy, Inc., for NASA, under contract NAS 5-26555.  Support for PFH was provided by the Miller Institute for Basic Research in Science, University of California Berkeley.  The computations in this paper were run on the Odyssey cluster supported by the FAS Research Computing Group at Harvard University.

\bibliographystyle{apj}
\bibliography{../smg}

\end{document}